\documentclass[twocolumn,showpacs,superscriptaddress,amsmath,amssymb]{revtex4}
\usepackage{graphicx}
\usepackage{dcolumn}
\usepackage{bm}

\def\bd{
\begin{document}} \def\ed{\end{document}}
\def\bmp{\begin{minipage}} \def\emp{\end{minipage}}
\def\bcc{\begin{center}} \def\ecc{\end{center}}     \def\npg{\newpage}
\def\beq{\begin{equation}} \def\eeq{\end{equation}} \def\hph{\hphantom}
\def\be{\begin{equation}} \def\ee{\end{equation}} \def\r#1{$^{[#1]}$}
\def\n{\noindent} \def\ni{\noindent} \def\pa{\parindent}
\def\hs{\hskip} \def\vs{\vskip} \def\hf{\hfill} \def\ej{\vfill\eject}
\def\cl{\centerline} \def\ob{\obeylines}  \def\ls{\leftskip}
\def\underbar#1{$\setbox0=\hbox{#1} \dp0=1.5pt \mathsurround=0pt
   \underline{\box0}$}   \def\ub{\underbar}    \def\ul{\underline}
\def\f{\left} \def\g{\right} \def\e{{\rm e}} \def\o{\over} \def\d{{\rm d}}
\def\vf{\varphi} \def\pl{\partial} \def\cov{{\rm cov}} \def\ch{{\rm ch}}
\def\la{\langle} \def\ra{\rangle} \def\EE{e$^+$e$^-$} \def\pt{p_{\rm t}}
\def\dt{\delta}   \def\sqnn{\sqrt{s_{\rm NN}}}
\def\bitz{\begin{itemize}} \def\eitz{\end{itemize}}
\def\btbl{\begin{tabular}} \def\etbl{\end{tabular}}
\def\btbb{\begin{tabbing}} \def\etbb{\end{tabbing}}
\def\beqar{\begin{eqnarray}} \def\eeqar{\end{eqnarray}}
\def\\{\hfill\break} \def\dit{\item{-}} \def\i{\item}
\def\bbb{\begin{thebibliography}{9} \itemsep=-1mm}
\def\ebb{\end{thebibliography}} \def\bb{\bibitem}
\def\bpic{\begin{picture}(260,240)} \def\epic{\end{picture}}
\def\akgt{\noindent{Acknowledgements}}
\def\fgn{\noindent{\bf\large\bf figure captions}}
\def\lan{\langle}
\def\ran{\rangle}
\def\p{\pi}
\def\ifmath#1{\relax\ifmmode #1\else $#1$\fi}%
\def\rc{\ifmath{{\mathrm{c}}}}
\def\cut{\ifmath{{\mathrm{cut}}}}
\def\rF{\ifmath{{\mathrm{F}}}}
\def\rK{\ifmath{{\mathrm{K}}}}
\def\rp{\ifmath{{\mathrm{p}}}}
\def\rt{\ifmath{{\mathrm{t}}}}
\def\LAB{\ifmath{{\mathrm{LAB}}}}
\def\cut{\ifmath{{\mathrm{cut}}}}
\def\beq{\begin{equation}}
\def\eeq{\end{equation}}
\def\us{^{(s)}}  \def\bea{\begin{eqnarray}} \def\eea{\end{eqnarray}}
\def\nbr{\nonumber} \def\e{\eta} \def\dt{\delta} \def\D{\Delta}
\def\r{\rho}
\newcommand{\cinst}[2]{$^{\mathrm{#1}}$~#2\par}
\newcommand{\crefi}[1]{$^{\mathrm{#1}}$}
\newcommand{\crefii}[2]{$^{\mathrm{#1,#2}}$}
\newcommand{\crefiii}[3]{$^{\mathrm{#1,#2,#3}}$}
\newcommand{\HRule}{\rule{0.5\linewidth}{0.5mm}}

\usepackage{color}
\newcommand{\Blue}[1]{\textcolor[named]{Blue}{#1}}
\newcommand{\blue}[1]{\textcolor[named]{Blue}{#1}}
\newcommand{\red}[1]{\textcolor[named]{Red}{#1}}
\newcommand{\violet}[1]{\textcolor[named]{Violet}{#1}}
\newcommand{\brown}[1]{\textcolor[named]{Brown}{#1}}
\newcommand{\green}[1]{\textcolor[named]{Green}{#1}}
\newcommand{\Red}[1]{\textcolor[named]{Red}{#1}}
\newcommand{\yellow}[1]{\textcolor[named]{Yellow}{#1}}
\newcommand{\magenta}[1]{\textcolor[named]{Magenta}{#1}}

\bd
\title{Is there hydrodynamic flow at RHIC ?}
\author{Wang Meijuan}
\affiliation{Institute of Particle Physics, Huazhong Normal
University, Wuhan 430079, China}  \author{Liu
Lianshou}\affiliation{Institute of Particle Physics, Huazhong
Normal University, Wuhan 430079, China}
 \affiliation{Key Laboratory of
Quak and Lepton Physics,  Ministry of Education of China }
 \author{ Wu Yuanfang} \affiliation{Institute of Particle Physics, Huazhong
Normal University, Wuhan 430079, China}
 \affiliation{Key Laboratory of
Quak and Lepton Physics,  Ministry of Education of China }

\begin{abstract}
It is argued that the observation of anisotropic azimuthal
distribution of final state particles alone is insufficient to
show whether the formed matter at RHIC behaves like hydrodynamic
flow. Examining the intrinsic interaction (or correlation) of the
formed matter should provide more definite judgement. To the end,
a spatial-dependent azimuthal multiplicity-correlation pattern is
suggested. It shows clearly in the pattern that there are two
kinds of interactions at the early stage of Au + Au collisions at
$\sqnn=200$ GeV generated by RQMD with hadron re-scattering and
AMPT with string melting. This is out of the expectation from the
elliptic flow driven by anisotropic expansion.
\end{abstract}


\maketitle

The data from current relativistic heavy ion experiments show that
a new form of matter --- quark-gluon plasma (QGP) has been
produced at RHIC~\cite{qgp, gulassys}. The second Fourier
coefficient $v_2$ of the anisotropic transverse-momentum $p_{\rm
T}$ distribution of final state particles is believed to provide
the anisotropic collective flow behavior at the early stage of
collision. The successful hydrodynamic description~\cite{mv2} on
the observed mass dependence of $v_2$ at $p_{\rm T} < 2$ GeV shows
that the observed dense matter behaves like a perfect fluid rather
than an ideal gas, and is, therefore, referred to as sQGP.

However, hydrodynamics can still not quantitatively fit the
observed mass dependence of $v_2$~\cite{qgp, break}. The recently
measured elliptic flow $v_2$ from Cu +Cu collisions at 200 GeV is
unexpected as large as that from Au + Au collisions at the same
energy~\cite{phenix,trainor}. Moreover, the resulting matter may
be treated as hydrodynamic flow only if the initial interaction
among the constituents are sufficiently strong to establish local
thermal equilibrium rapidly, and then to maintain it over a
significant evolution time. No known strong-interaction process
could be thermalized on such a short timescale. A Liquid without
viscosity is also hard to be understood
theoretically~\cite{viscosity}. So, there appear a number of
alternative non-equilibrium treatments, which have also been
compared to RHIC data~\cite{bmuller,rudy,yezhov}.

To conclusively clarify the debate, a direct experimental
examination on the intrinsic interaction of the formed matter is
neccessary. The hydrodynamic flow at RHIC is supposed to be driven
by the so called anisotropic expansion. In non-central collisions,
the initial participant zone of the two colliding nuclei is
approximately an ellipse, and the density gradient along the short
side of ellipse is larger than that along the long side. It is
argued that the larger density, or pressure, gradient along the
short side of ellipse makes collective expansion to be privileged
in this direction, i.e., the anisotropic expansion, producing
in-plane elliptic flow, or the transverse-momentum of final state
particles distribute in an ellipse perpendicular to the one in
coordinate space.

However, the main physical quantity which can be extracted from
experimental data in exploiting relativistic hydrodynamic approach
is the elliptic flow parameter $v_2$. It only indicates the
possible preferential direction of expansion and contains no
information on the intrinsic interaction of the formed matter. It
therefore is insufficient to assure whether the formed matter
behaves like hydrodynamic flow.

If the anisotropic expansion is the only driver of the elliptic
flow, the distribution of intrinsic interaction (or correlation)
of flow should have the same anisotropy, i.e., in-plane like.
Moreover, if it is really hydrodynamic flow, it should be well
locally thermalized and reach thermal equilibrium. Then all other
interaction history before anisotropic expansion should be
forgotten. These characteristics can be examined in an
experimentally measurable correlation pattern.

In this letter, we will first introduce the spatially-dependent
correlation pattern, i.e., neighboring angular-bin multiplicity
correlation pattern. Then, we demonstrate that at least two kind
of interactions are revealed by the suggested correlation pattern
in Au + Au collisions at 200GeV, generated by RQMD~\cite{rqmd} and
AMPT~\cite{ampt}. Finally, how to experimentally measure the
correlation pattern and anisotropic correlation coefficient is
discussed.

To examine the intrinsic interaction of highly anisotropic system,
a spatial-dependent bin-bin correlation is called for.
Conventionally, the spatially averaged bin-bin correlation has
been used in multiparticle production in exploring self-similar
fractality~\cite{bialas}, where the system is supposed to be
homogeneous, and only scaling in the shrinking of phase space is
concerned. Another intrinsic interaction related measure is the
2-particle azimuthal correlation~\cite{2par,phenix}. It concerns
the average correlation of two particles separated by a certain
angle, no matter where the two particles are in the azimuthal
space. It therefore can not tell us where the preferential
direction of intrinsic interactions are.

The newly suggested {\it spatial-dependent} neighboring bin
correlation pattern~\cite{wu-pre} provides a typical spatial
distribution of two-bin correlation. The information on intrinsic
correlation can be well presented by the measure, and it should
give more direct and definite judgement on the properpty of the
formed matter at RHIC.

It is well-known that the general 2-bin correlation is defined as
\beq C_{m_1,m_2}= \frac{\langle n_{m_1} n_{m_2} \rangle}{\langle
n_{m_1} \rangle \langle n_{m_2} \rangle}-1, \eeq \noindent where
$m_1$ and $m_2$ are the positions of the two bins in phase space
and $n_m$ is the measured content in the $m$th bin.

We divide the $2\pi$ azimuthal angle equally into $M$ bins and
specify $n_m$ as the multiplicity in the $m$th angular bin. If we
let $m_1=m$ and $m_2=m+1$, $C_{m_1,m_2}$ is reduced to the {\it
neighboring angular-bin multiplicity correlation pattern}, \beq
C_{m,m+1}=\frac{\langle n_{m}n_{m+1}\rangle}{\langle
n_{m}\rangle\langle n_{m+1}\rangle}-1. \eeq \noindent  It is clear
that the correlation pattern measures how the nearby particles
correlate with each other in different directions of azimuthal
space. If the particles are produced independently in the whole
phase space, then $\langle n_mn_{m+1}\rangle=\langle
n_m\rangle\langle n_{m+1}\rangle$, and $C_{m_1,m_2}$ vanishes.

In order to apply this correlation pattern to current relativistic
heavy ion collision, we choose the RQMD and AMPT models as
examples. The RQMD (relativistic quantum molecular dynamics) with
re-scattering is a hadron-based transport model~\cite{rqmd}. The
final hadron interactions are implemented in the model by hadron
re-scattering. The anisotropic collective flow produced by the
model is much smaller than the observed data at RHIC. In contrary
to the RQMD model, the AMPT is a multi-phase transport model,
where both hadron and parton interactions are taken into account.
In the AMPT with string melting, the parton level transport is
fully taken into account, and the observed anisotropic collective
flow at RHIC is well reproduced~\cite{ampt}.

For Au + Au collisions at $\sqnn=200$ GeV, we generate 249,824 and
204,004 events using RQMD with hadron re-scattering and AMPT with
string melting, respectively. Their neighboring angular-bin
multiplicity correlation patterns are shown in Fig.~1(a) by open
and solid circles, respectively. Here we partition the whole
azimuthal range $2\pi$ uniformly into 50 equal size angular bins.
$\phi =0$ refers to the direction of the reaction plane in nuclear
collision. The errors are statistical only and most of them are
smaller than the symbol size in this and following figures. It is
clearly shown in Fig.~1(a) that correlation patterns from these
two models are $-\cos2\phi$ (out-of-plane) like, opposite to the
well-known $\cos2\phi$ (in-plane) liked azimuthal distribution.
This is in contrary to the expectation that the formed matter
expands collectively toward in-plane direction. Some unexpected
interactions should be responsible for such a result.

In order to see how the results come, the centrality dependence of
neighboring angular-bin multiplicity correlation patterns from
these two models are presented in Fig.~1(b) and (c), respectively,
where three typical centralities are specified in the legends. One
can observe that the two models give qualitatively the same
centrality dependence of azimuthal correlation pattern. The
correlation patterns are $\cos2\phi$ like in peripheral
collisions, then turn to flat in mid-central collisions, and
become $-\cos2\phi$ like in near-central collisions. Here, we
present only three centrality ranges to show their typical
behavior. In fact, the correlation pattern changes gradually from
$\cos2\phi$ to $-\cos2\phi$ with centrality. It is clear that two
opposite trends dominate in peripheral and near-central
collisions, respectively. In the mid-central collisions, the two
trends turn to balance and the correlations become equal in all
directions. Moreover, these characteristics are independent of the
specific assumptions implemented in the two models, in particular
independent of the hadronization schemes assumed in the models.

\begin{figure}
\includegraphics[width=3.4in]{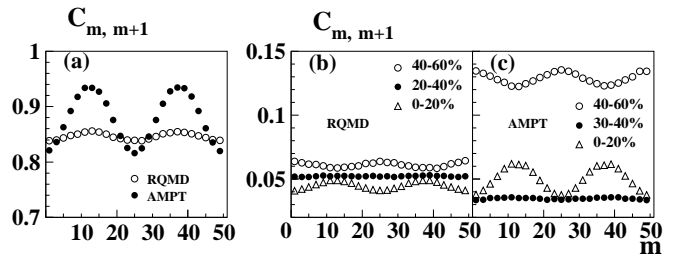}
\caption{\label{Fig. 1} (a)The neighboring angular-bin correlation
patterns for Au + Au collisions at 200 GeV from the RQMD with
re-scattering and AMPT with string melting. The centrality
dependence of the correlation patterns from (b)the RQMD with
re-scattering , and (c) AMPT with string melting.}
\end{figure}

The characteristics of correlation pattern reveal that there are
two opposite intrinsic interactions in the formed matter in these
two transport models. One has the same preferential direction as
the anisotropic expansion. The other one is opposite to it. This
also shows that the anisotropic azimuthal distribution is not only
driven by anisotropic expansion. It is resulted from the
combination of these two opposite interactions.

The anisotropic expansion and the late hadronization are
impossible to produce strong correlations in out-of-plane
direction. Only the initial source eccentricity in non-central
collisions is preferential in the direction. It results in a
larger initial number of participant nucleons in the out-of-plane
direction, which in turn could generate stronger interaction in
the direction. As long as the system is not fully thermalized,
this initial interaction will compete with the subsequent
anisotropic expansion.

In peripheral collisions, the overlap zone is small and so is the
number of participant nucleons, but the difference between the
minor and major axes of overlap ellipse is large, and so is the
difference of pressure gradients. In this case the anisotropic
expansion dominates the final observables, and the effects of
initial interaction in correlation patterns are hidden. In
near-central collisions, the overlap zone becomes large and the
difference between minor and major axes of ellipse is small, so
that the initial interactions are strong enough to show themselves
up in final observable. This is why the out-of-plane correlation
patterns appear at near-central collisions.

So the behavior of the formed matter in these two transport models
are far from the flow in relativistic hydrodynamics sense. This is
out of the current expectation for the formed matter at RHIC.
Measuring the correlation pattern by the data of relativistic
heavy ion collisions at RHIC and LHC is therefore looking forward.
As long as the observed preferential direction of correlation
pattern are different from that of its azimuthal distribution,
such as what we show by transport models, then there should be no
hydrodynamic flow at RHIC. On the other hand, if the
experimentally measured correlation pattern has the same
anisotropy as its azimuthal distribution, it will be a strong
support to the current expectation at intrinsic interaction level
that the formed matter at RHIC indeed behaves like hydrodynamic
flow.

The correlation patterns are typical periodic functions of
azimuthal angle in peripheral and central collisions, as shown in
Fig.~1. So they can be well expanded by Fourier series, \beq
C_{\phi,\phi+\delta\phi}= C \f[ 1+\sum_{i=1}
2u_i\cos(i(\phi-\psi_r)) \g], \eeq where the $\psi_r$ is the
direction of reaction plane, and is zero in the model analysis.
But in real experimental data analysis, it has to be determined
event-by-event, and thereby refers to event-plane. It has been
carefully estimated in the measurement of anisotropic elliptic
flow $v_2$ in current relativistic heavy
experiments~\cite{phi-r,starflow}. The main contribution in the
expansion series comes from $\cos2(\phi-\psi_r)$. Its coefficient
$u_2$ provides the preferential direction and strength of
anisotropic correlation pattern. We specify it as {\it anisotropic
correlation coefficient} (ACC). It will make the systematic study
of the correlation pattern easy.

\begin{figure}
\includegraphics[width=3.4in]{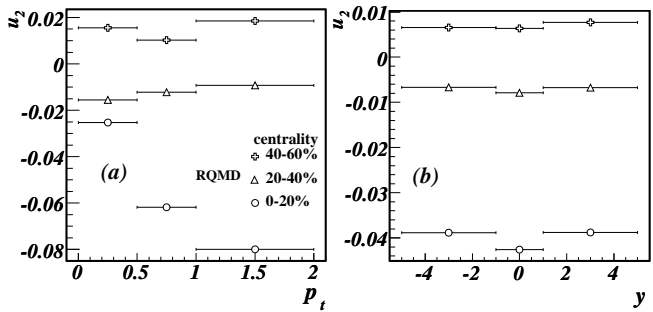}
\caption{\label{Fig. 2} (a) Transverse-momentum and (b) rapidity
dependence of $u_2$ at three centralities for Au + Au collision at
200GeV from RQMD with re-scattering.}
\end{figure}

It is interesting to see how ACC, $u_2$, depends on the
transverse-momentum $p_t$ of final state particles, in comparison
to the corresponding $p_t$ dependence of $v_2$. It is known that
the evolution schemes of RQMD with re-scattering and AMPT with
string are different. In the former, the $p_t$ spectrum is
determined by the temperature of thermal source. High $p_t$
particles are emitted early at high temperature and low $p_t$ ones
are emitted later on at low temperature. The range of $p_t$ of
final state particles is related to its emitting
proper-time~\cite{rqmd}. So the $p_t$ dependence of $u_2$ in RQMD
with re-scattering will present how the correlation pattern
changes with evolution.

The results are presented in Fig.~2(a). We can see that in each
$p_t$ interval, $u_2$ keeps positive in peripheral collisions,
becomes negative for mid-central collisions, and becomes even more
negative for central collisions. They are similar to that for all
$p_t$ particles shown in Fig.~1(b). It should also be noticed in
Fig.~2(a) that $u_2$ is almost independent of the choice of $p_t$
ranges of final state particles in peripheral and mid-central
collisions, but decrease rapidly with the increase of $p_t$ in
central collision. This is understandable since high $p_t$
particles are emitted earlier, and less influenced by the later
anisotropic expansion in central collisions.

On the contrary, each parton in the AMPT with string melting has
its own freeze-out time, which span a long period after the
initial interaction of the two nuclei, and are unrelated to each
parton's transverse momentum~\cite{liu-yu, ampt}. So similar $p_t$
dependence of $u_2$ can be observed in this model only when the
chosen interval of $p_t$ is very large.

The rapidity dependence of ACC, $u_2$, is further studied by these
two transport models. They give qulitatively the same dependency.
The results from RQMD are presented in Fig.~2(b), where three
typical rapidity ranges, i.e., forward, backward and central
rapidity ranges, are chosen. It shows that the correlation pattern
is independent of the choice of rapidity range, and similar to
that in the whole rapidity space. So in finite rapidity ranges of
current relativistic heavy ion experiments~\cite{starflow},
studying the correlation pattern and anisotropic correlation
coefficient is expectable, and will provide more definite evidence
on whether or not the formed matter behaves like hydrodynamic
flow.

To the summary, it is argued that the observation of anisotropic
azimuthal distribution of final state particles alone is
insufficient to assure whether the formed matter at RHIC behaves
like hydrodynamic flow. Examining the intrinsic interaction (or
correlation) of the formed matter should provide more definite
judgement. To the end, a spatially-dependent azimuthal
multiplicity-correlation pattern is suggested. It shows clearly
that there are two kinds of interactions at early stage of Au + Au
collisions at $\sqnn=200$ GeV, generated by RQMD with hadron
re-scattering and AMPT with string melting. One is in-plane
preferential as expected from anisotropic expansion due to initial
eccentricity in non-central collisions. Another new one is
out-of-plane preferential, which may be resulted from the larger
initial number of participant nucleons in these direction. These
characters of correlation pattern show at least in two transport
models that the formed matter does not behave like hydrodynamic
flow, in contrary to current expectation. Finally, how to
experimentally measure the correlation pattern and anisotropic
correlation coefficient is discussed.

The authers would thank Dr. Nu Xu, Aihong Tang and Huangzhong Huan
for their stimulating comments. We are grateful for the financial
supports from the NSFC of China under projects: No. 90503001,
10610285, 10775056.

\ed